\documentclass[aps,prx,twocolumn,letterpaper,superscriptaddress]{revtex4-2}
\usepackage{graphicx}
\usepackage{amsmath}
\usepackage{amssymb}
\usepackage{url}
\usepackage{hyperref}

\begin{document}
\title{On the infrared limit of the\\
$\mathrm{O}(3)$ nonlinear $\sigma$-model at $\theta = \pi$}
\author{Martin R. Zirnbauer}
\affiliation{Institut f\"ur Theoretische Physik, Universit\"at zu K\"oln, Z\"ulpicher Str. 77a, 50937 K\"oln, Germany}
\date{\today}

\begin{abstract}
2D nonlinear $\sigma$-models with Hermitian symmetric target admit a $\theta$-term, which couples the field theory to the topological charge of its instanton gas. At the special coupling $\theta = \pi$, by what is nowadays attributed to a coupling-constant anomaly of Lieb-Schultz-Mattis type, such models have a degenerate ground state. Yet, the details of their non-trivial infrared limit have remained open in general. Here we suggest that non-perturbative renormalization group flow into the strong-coupling regime induces strong fluctuations of the $\theta$-parameter, with the consequence that the instanton density is suppressed, the target-space topology effectively altered, and the target-space metric driven off reality and into geometrostasis. Assuming this heuristic scenario and combining it with a Cauchy process of target-space deformation, we present a detailed argument that the $\mathrm{O}(3)$ nonlinear $\sigma$-model at $\theta = \pi$, known to be the effective field theory for critical antiferromagnetic quantum spin chains with large half-integer spin, renormalizes to the conformal field theory of a $\mathrm{U}(1)$ boson with compactification radius $r = 1 / \sqrt{2}$. A closely related scenario applies to Pruisken's nonlinear $\sigma$-model for the integer quantum Hall transition.
\end{abstract}
\maketitle

\section{Introduction}
\label{sect:intro}

Nonlinear $\sigma$-models are field theories of mappings into a Riemannian symmetric space $G/K$. In their lower critical dimension of $d_c = 2$, the case of interest here, they have a dimensionless coupling and are scale-invariant in the classical limit. Subject to regularization and renormalization as quantum field theories, they typically undergo dynamical mass generation when the symmetry group $G$ is compact and non-Abelian. This scenario, closely parallel to the one for Yang-Mills theories in four dimensions, is known as the \emph{mass-gap conjecture}.

Exceptions to the mass-gap rule are provided by 2D nonlinear $\sigma$-models with a so-called $\theta$-term. These models have a target space of Hermitian type (i.e., one from $\mathrm{U} / \mathrm{U}\! \times \! \mathrm{U}$, $\mathrm{Sp} / \mathrm{U}$, or $\mathrm{O} / \mathrm{U}$), carrying a $G$-invariant and closed two-form $\omega$ related to the metric tensor of the Riemannian geometry of $G/K$. The  existence of $\omega$ (representing a Chern class) is accompanied by field configurations with instantons. As the topological coupling, an angle $\theta$, is tuned through the symmetry-enhanced point $\theta = \pi$, the model undergoes a phase transition, which is known in some cases to be of second order, i.e.\ with a divergent correlation length and massless excitations.

A prominent example in the latter category is the so-called $\mathrm{O}(3)$ nonlinear $\sigma$-model at $\theta = \pi$ (or $\mathrm{SM}_{\pi}$ for short). Following the pioneering work of \cite{PW84} and \cite{ZZ92}, this model is known to be massless. One therefore expects it to flow under the renormalization semigroup (RG) to a fixed point with conformal symmetry describing the physical observables in the infrared limit.

Given that scenario, one asks about the precise nature of the RG-fixed point theory. The standard answer \cite{AH87} to that question is the Wess-Zumino-Novikov-Witten (WZW) model for the group $\mathrm{SU}(2)$ [as the universal cover of $\mathrm{SO}(3) \subset \mathrm{O}(3)$] with current-algebra level $k = 1$. Here it should be noted that the surmised RG flow from $\mathrm{SM}_\pi$ to the $\mathrm{SU}(2)_{k=1}$ WZW model implies a large degree of symmetry enhancement, from the global $\mathrm{O}(3)$ symmetry of the former to the infinite-dimensional affine symmetry $\mathfrak{su}(2)_L \times \mathfrak{su}(2)_R$ of the latter. That mechanism of symmetry enhancement has been found to be very fragile.

The puzzle motivating the present paper can now be stated as follows. Among the set of models with superspace target, there exists a number of 2D nonlinear $\sigma$-models at $\theta = \pi$ which defy the mass-gap paradigm by remaining scale-invariant in the infrared. As effective field theories, these models are meant to describe the critical behavior at Anderson-type phase transitions between disordered free-fermion topological insulators and superconductors in two dimensions. (In the language of the Tenfold Way \cite{HHZ05}, they are referred to by their Cartan types as $A$, $C$, $D$, $A{\rm I\!I}$, $D{\rm I\!I\!I}$; a prominent example is the Pruisken model for the integer quantum Hall transition.) None of them, however, is a plausible candidate for a fixed point of the renormalization group flow. At the same time, the mentioned scenario of symmetry enhancement $G \to \mathfrak{g}_L \times \mathfrak{g}_R$ is generically unstable with respect to perturbations preserving criticality and the global $G$-symmetry; by that token, it can be considered as ruled out for these models \cite{RS01, CFT-IQHT} (as well as many others \cite{Affleck88}).

The puzzle then is this: it appears that the scenario of symmetry enhancement $G \to \mathfrak{g}_L \times \mathfrak{g}_R$ is the \emph{only} mechanism (assuming the setting of a non-Abelian symmetry group $G$) which is capable of stopping the RG flow to produce an infrared fixed point \cite{BCZ85}. If that was really the case, how should we think about and analyze those topological quantum phase transitions of Anderson type in two dimensions? What are the renormalized field theories describing their infrared physics? The severity of the puzzle is highlighted by recent claims \cite{KCGM21, KGM22} that the principle of conformal symmetry is violated for them.

In view of that conundrum, we are intrigued by the observation \cite{Ginsparg88} that the conformal field theory of the $\mathrm{SU} (2)_1$ WZW model has an equivalent description by a $\mathrm{U}(1)$ boson at compactification radius $r = 1 / \sqrt{2}$ (using string theory conventions) and with hidden $\mathrm{SU}(2)$ symmetry. This observation prompts two questions:
\begin{enumerate}
\item[(i)] Can one draw up a plausible scenario of renormalization group flow from the $\mathrm{O}(3)$ nonlinear $\sigma$-model at $\theta = \pi$ to that $\mathrm{U}(1)_{r = 1/\sqrt{2}}$ boson model?
\item[(ii)] If so, does that scenario admit generalization to the uncharted territory of super-target nonlinear $\sigma$-models still looking for their RG-fixed points?
\end{enumerate}
While we firmly believe the answer to be positive for both questions, the present paper will address the first one only. (As for the second one, let us announce that the scenario to be developed here does extend to that of \cite{CFT-IQHT}, where a conformal field theory was proposed for the infrared limit of the integer quantum Hall transition.)

To prove our answer to question (i), we would need to carry out the full program of non-perturbative renormalization. While that is beyond our current abilities, we shall offer a fairly detailed heuristic picture. In brief, we suggest that the axis of the coarse-grained $\sigma$-model field settles down, for any configuration of statistical relevance, in some ``easy plane'' $\mathrm{S}^1 \subset \mathrm{S}^2$ (instead of roaming throughout the suspension $\mathrm{SU}(2) \equiv \mathrm{S}^3 \supset \mathrm{S}^2$), thereby avoiding the fate of mass gap and exponential decay of correlations. Small fluctuations around that easy plane induce the fixed-point stiffness of its $\mathrm{U}(1) = \mathrm{S}^1$ field.

In a little more detail, our heuristic picture goes as follows. The RG flow into the strong-coupling regime is expected to generate $\mathrm{O}(3)$-orbit variations and hence strong fluctuations of the $\theta$-parameter. Consequently, by the Fourier duality between $\theta$ and the topological density, instantons get suppressed on large length scales. So, as the RG flow reaches the infrared, the target-space topology is effectively altered from spherical ($\mathrm{S}^2$) to cylindrical ($\mathrm{S}^1 \times \mathbb{R}$), for any field configuration of statistical significance. That change makes it possible to deform the $\sigma$-model, by holomorphic continuation of its $\mathrm{O}(3)$-invariant geometry, to a nonlinear model with complex metric structure on the target-space cylinder $\mathrm{S}^1 \times \mathbb{R}$. The latter features an $\mathbb{R}$-valued Gaussian field, which can be integrated out to produce an effective action for the $\mathrm{S}^1$-valued field. In this way, assuming continued RG flow to strong coupling, we actually arrive at the $\mathrm{U}(1)_{r = 1/\sqrt{2}}$ conformal field theory with hidden $\mathrm{SU}(2)$ symmetry.

The contents of the present paper are as follows. In Sect.\ \ref{sect:O3} we define the $\mathrm{O}(3)$ nonlinear $\sigma$-model at $\theta = \pi$ (abbreviated as $\mathrm{SM}_\pi$). We mention its role as an effective field theory for critical antiferromagnetic quantum spin chains with large half-integer spin (\ref{sect:H-AF}). We review the reasons why $\mathrm{SM}_\pi$ is considered to be massless (\ref{sect:LSM}). We also review the $\mathrm{SU}(2)_{k=1}$ WZW model as the possible end point of a renormalization-group trajectory starting at $\mathrm{SM}_\pi$ ({\ref{sect:WZW}). In Sect.\ \ref{sect:U(1)} we define the $\mathrm{U}(1)_{r = 1/\sqrt{2}}$ conformal field theory. We extend the $\mathrm{U}(1) = \mathrm{S}^1$ target space to a cylinder, $\mathrm{S}^1 \times \mathbb{R}$, introducing a real-valued field for the second factor, in order to make $\mathrm{SU}(2)$ manifest as an infinitesimal symmetry (\ref{sect:extend}). We compute the two-loop RG beta function of the extended theory to demonstrate perturbative one-parameter renormalizability (\ref{sect:Ricci}).

In a substantial section, \ref{sect:heuRG}, we develop a heuristic RG scenario for $\mathrm{SM}_\pi$. We review what is known from perturbative renormalization (\ref{sect:RG-pert}). Turning to real-space non-perturbative renormalization
(\ref{sect:RG-thoughts}), we show that a good choice of averaging map for the Kadanoff block-spin transformation gives a direct intuition for the workings of perturbative RG (\ref{sect:arithmean}). Armed with that insight, we argue that the coupling ``constants'' of $\mathrm{SM}_\pi$ become fluctuating variables on entry into the strong-coupling regime (\ref{sect:punch}). Our main hypothesis then is that fluctuations of the $\theta$-parameter act to suppress the topological density (\ref{sect:fate-theta}). That serves as the justification for target-space surgery and reconstruction (\ref{sect:surgery}), $\mathrm{S}^2 \to \mathrm{S}^1 \times \mathbb{R}$. In Sect.\ \ref{sect:cauchy} we exploit the proposed change of target-space topology to carry out an exact deformation from $\mathrm{SM}_\pi$ to the $\mathrm{S}^1 \times \mathbb{R}$ extension of the $\mathrm{U}(1)_{r = 1 / \sqrt{2}}$ conformal field theory. Sect.\ \ref{sect:sum} contains a summary and concise guide to our main thread of thought. The issue of symmetry under finite (as opposed to infinitesimal) $\mathrm{O}(3)$ transformations is also addressed there.

\section{$\mathrm{O}(3)$ nonlinear $\sigma$-model}
\label{sect:O3}

The $\mathrm{O}(3)$ nonlinear $\sigma$-model is a field-theoretical model for a real three-component vector field, say
\begin{equation}
    x \mapsto \vec{m}(x) = \big( m^1(x), m^2(x), m^3(x) \big) ,
\end{equation}
constrained to be of unit length, $\vec{m} \cdot \vec{m} = 1$. Thus $\vec{m}(x)$ takes values in the unit sphere $\mathrm{S}^2 \subset \mathbb{R}^3$. The model's name reflects its symmetry under global $\mathrm{O}(3)$ transformations; cf.\ below. The continuum version of the model in two dimensions is given by a functional integral
\begin{equation}\label{eq:Z-NLsM}
    \begin{split}
    Z_{\beta,\theta}^{\mathrm{O}(3)} &= \int \mathcal{D} m \; \mathrm{e}^{- S_{\beta,\theta}[\vec{m}]} \,, \cr S_{\beta,\theta} [\vec{m}] &= \int_\Sigma d^2x \left(  \frac{\beta}{8\pi} \, \partial^\mu \vec{m} \cdot \partial_\mu \vec{m} + \frac{\mathrm{i} \theta}{2\pi}  \mathcal{L}_{\rm top} \right) , \\ \mathcal{L}_{\rm top} &= \epsilon^{\mu\nu}\, {\textstyle{\frac{1}{4}}} \vec{m} \cdot (\partial_\mu \vec{m} \times \partial_\nu \vec{m}),
    \end{split}
\end{equation}
with (informally stated) path-integral measure
\begin{equation}\label{eq:meas}
    \mathcal{D} m = \prod\nolimits_x \omega_x
\end{equation}
where $\omega = \frac{1}{4} \epsilon_{abc} \, m^a dm^b \wedge dm^c$ normalized by $\int_{\mathrm{S}^2} \omega = 2\pi$ is half the standard solid-angle two-form on $\mathrm{S}^2$.

The parameters of the theory are the metric coupling $\beta$ (inverse temperature, or spin stiffness) and the coupling $\theta$ multiplying the topological density function $\mathcal{L}_{\rm top}$. The symbol $\epsilon^{\mu\nu}$ stands for the epsilon tensor expressing the area two-form of the oriented two-dimensional space (or space-time), $\Sigma$, and indices are raised, $\partial^\mu = \delta^{\mu\nu} \partial_\nu$, via the metric tensor $\delta^{\mu\nu}$ of $\Sigma$. Note that the metric term with coupling $\beta$ is real-valued, while the $\theta$-term takes values in the imaginary numbers. For a closed  space(-time) manifold $\Sigma$, the parameter $\theta$ has the mathematical meaning of an angle as $\exp(- S_{\beta,\theta}) = \exp(- S_{\beta,\theta + 2\pi})$ by quantization of the topological charge,
\begin{equation}\label{eq:topo-q}
    q \equiv \frac{1}{2\pi} \int_\Sigma d^2 x \, \mathcal{L}_{\rm top} \in \mathbb{Z} .
\end{equation}
In the following we take $\Sigma$ to be the Euclidean plane $\mathbb{R}^2$ with Euclidean metric $\delta^{\mu\nu}$ (although a better choice for rigorous work would be $\Sigma = \mathrm{S}^2$).

The symmetry group $\mathrm{O}(3)$ of the model acts by global transformations,
\begin{equation}\label{eq:O3-sym}
    m^a(x) \mapsto R_{\; b}^a \, m^b(x), \quad  R \in \mathrm{O}(3) .
\end{equation}
These are symmetries of the action functional (and the path-integral measure) when $R \in \mathrm{SO}(3)$. For $R \in \mathrm{O}(3)$ with $\mathrm{Det}(R) = - 1$, the transformation (\ref{eq:O3-sym}) reverses the sign of the solid-angle two-form $\omega$ and hence the sign of the topological coupling $\theta$; therefore, such $R$ are symmetries only when combined with orientation reversal (also known as a parity transformation) of $\Sigma$.

Our interest in the following is in the model for $\theta = \pi$. In that special case, orientation reversal of $\Sigma$ (or sign reversal $\theta \to -\theta$) is a symmetry by the equivalence $\theta \sim \theta + 2\pi$ due to quantization of the topological charge ($q \in \mathbb{Z}$) for $\Sigma$ without boundary. This symmetry persists under renormalization and thus constrains the possible scenarios for the infrared limit of the theory.

\subsection{Motivation: antiferromagnetic spin chains} \label{sect:H-AF}

There exist various motivations to study the $\mathrm{O}(3)$ nonlinear $\sigma$-model as a model of field theory and statistical mechanics. Our primary motivation here is its justification as an effective field theory (or continuum description at long wavelengths) for anti-ferromagnetic (AF) quantum spin chains with space translation symmetry. The latter are 1D quantum systems with Hamiltonian
\begin{equation}\label{eq:H-AF}
    H = J \sum_{{n} \in \mathbb{Z}} \; \sum_{a=1}^3 S_n^a S_{n+1}^a \quad (J > 0) ,
\end{equation}
acting on the direct-product Hilbert space $\otimes_n \mathbb{C}^{2|S|+1}$ for a chain of sites (labeled by the integers, $n \in \mathbb{Z}$), each carrying the spin-$|S|$ representation of the rotation group. The spin operators $S_n^a$ ($a = 1, 2, 3$) represent the Lie algebra of rotation generators, i.e.,
\begin{equation}
    [\mathrm{i} S_n^a , \mathrm{i} S_{n^\prime}^b] = \delta_{n n^\prime} f_c^{ab} \, \mathrm{i} S_n^c \,,
\end{equation}
with $f_c^{ab} = - \epsilon^{abc}$ the structure constants of $\mathrm{Lie}\; \mathrm{O}(3)$. Similar to Eqs.\ (\ref{eq:Z-NLsM}, \ref{eq:O3-sym}) the Hamiltonian (\ref{eq:H-AF}) is invariant under global rotations $S_n^a \mapsto R_{\; b}^a \, S_n^b$ by $R \in \mathrm{O}(3)$.

It is a classic fact (see, e.g.\ \cite{Fradkin-book}) that the action functional (\ref{eq:Z-NLsM}) can be derived from the Hamiltonian (\ref{eq:H-AF}) by the path-integral method with spin-coherent states. The derivation (by expansion in $1/|S|$) is controlled in the semiclassical limit of large spin $|S|$, and the result for the couplings is
\begin{equation}\label{eq:couplings}
    \beta = 2\pi |S| = \theta .
\end{equation}
Thus one has $\theta = 0$ (modulo $2\pi$) for the case of integer spin $|S|$, and $\theta = \pi$ (mod $2\pi$) for $|S|$ half-integer. Here it should be noted that the spin-$|S|$ representations for half-integer $|S|$ are projective (i.e.\ fail to be single-valued). By Eq.\ (\ref{eq:couplings}), this anomalous feature is passed on to the effective field theory at $\theta = \pi$.

\subsection{$\mathrm{SM}_\pi$ is massless}\label{sect:LSM}

A brief review of pertinent spin-chain phenomenology is as follows. For integer spin $|S|$, by what became known as Haldane's conjecture \cite{FDMH83}, the quantum spin chain (\ref{eq:H-AF}) is expected to have a unique ground state with a finite energy gap for excitations and hence exponential decay of spin-spin correlations. The prime motivation for that conjecture came from field theory: in weak-coupling perturbation theory, the stiffness parameter $\beta$ of the nonlinear $\sigma$-model decreases under renormalization, which implies that the spin chain is driven to an ``atomic'' limit. If so, the $\mathrm{O}(3)$ global symmetry becomes trivially represented in the infrared limit, for $\theta / 2\pi = |S| \in \mathbb{N}$.

In contrast, for translation-invariant chains with half-integer spin ($|S| \in \mathbb{N} + 1/2$) the low-energy behavior is qualitatively different. According to the Lieb-Schultz-Mattis theorem \cite{LSM61} and later work by Affleck and Lieb \cite{AL86}, the ground state in that case acquires a degeneracy which is at least two-fold in the limit of infinite chain length. It is then natural to expect that these spin chains, and hence the $\mathrm{O}(3)$ nonlinear $\sigma$-model at $\theta = \pi$ [or $\mathrm{SM}_{\pi}$ for short], are gapless with algebraically decaying correlations. Here a milestone was the work of Polyakov and Wiegmann \cite{PW84}, who reformulated $\mathrm{SM}_{\pi}$ as a Bethe-ansatz integrable interacting fermion model with infinitely many flavors. For the latter, they computed the free energy in an external field to demonstrate that the model has scale invariance and hence gapless (or massless) excitations. The massless scenario was later underpinned by Zamolodchikov and Zamolodchikov \cite{ZZ92}. These authors conjecturally attributed to $\mathrm{SM}_\pi$ an $S$-matrix with factorized scattering and an emergent symmetry $\mathrm{SU}(2)_L \times \mathrm{SU}(2)_R$ in the infrared limit.

Taking the modern perspective of symmetry-protected topological (SPT) phases, one may argue that the models at $\theta = 0$ and $\theta = 2\pi$ actually represent different SPT phases, as can be seen by considering a space-time $\Sigma$ with boundary, or by coupling the model to a background gauge field so as to exhibit an anomaly in the space of coupling constants \cite{CFLS20}. It follows that a phase transition must occur somewhere in the interval $0 < \theta < 2\pi$; on the symmetry grounds stated earlier, that transition is at $\theta = \pi$. In view of the results of \cite{PW84, ZZ92}, the transition point has massless excitations, and the $\mathrm{O}(3)$ global symmetry is non-trivially represented in the infrared limit.

All this begs the question: what exactly happens to $\mathrm{SM}_{\pi}$ under renormalization? This question has a widely accepted answer, which we review in the next subsection. (In the sequel, we shall offer another answer.)

\subsection{$\mathrm{SU}(2)_{k=1}$ WZW model}\label{sect:WZW}

Following \cite{PW84, ZZ92} and the work by Affleck and Haldane \cite{AH87}, it is believed \cite{CFLS20} that the $\mathrm{O}(3)$ nonlinear $\sigma$-model at $\theta = \pi$ renormalizes to the Wess-Zumino-Novikov-Witten (WZW) model for the group $\mathrm{SU}(2)$ with current-algebra level $k = 1$. Operationally, that scenario can be succinctly described as follows. From weak-coupling perturbation theory of the $\sigma$-model, we know that large values of $\beta$ decrease under renormalization.
As the strong-coupling regime of small $\beta$ is entered, the coupling ``constants'' $\beta$ and $\theta$ are expected to develop correlated fluctuations and become spatially inhomogeneous (see Sect.\ \ref{sect:RG-thoughts}). To account for these fluctuations, one may parameterize them in terms of an emergent dynamical field, say $x \mapsto \psi(x) \in [0,\pi]$. This prompts the idea of an \emph{interpolating} field theory where one evaluates the $\sigma$-model partition function (\ref{eq:Z-NLsM}) in a background of variable couplings
\begin{equation*}
    \beta(x) \equiv \beta \big(\psi(x) \big) , \quad \theta(x) \equiv \theta \big( \psi(x) \big)
\end{equation*}
and then post-averages the outcome with a sine-Gordon measure for $\psi$:
\begin{equation}\label{eq:Z-inter}
    \begin{split}
    &Z^\prime = \int \prod_x \sin^2 \psi(x) \, d\psi(x) \; \mathrm{e}^{- S_{\rm SG}[\psi]} Z_{\beta(\psi),\theta(\psi)}^{\mathrm{O}(3)} \,, \cr
    &S_{\rm SG} = \frac{k}{4\pi} \int_\Sigma d^2 x\, \partial^\mu \psi \, \partial_\mu \psi + M^2 \int_\Sigma d^2 x \, \cos^2 \psi \,.
    \end{split}
\end{equation}
From here, the WZW model (for $\mathrm{SU}(2)$ at level $k \in \mathbb{N}$) is obtained by setting $M^2 = 0$ and
\begin{equation}\label{eq:dynamo}
    \begin{split} \beta(\psi(x)) &= 2k\, \sin^2 \psi(x) \,, \cr \theta(\psi(x)) &= 2k \, \psi(x) - k \sin \, 2\psi(x) . \end{split}
\end{equation}
Geometrically speaking, $\psi$ is the radial variable for the suspension of $\mathrm{S}^2$ into $\mathrm{S}^3 \cong \mathrm{SU}(2)$. More explicitly, the emerging WZW field $x \mapsto g(x) \in \mathrm{SU}(2)$ is given by
\begin{equation}
    g = \mathbf{1} \cdot \cos\psi + \mathrm{i} \sigma_a m^a \sin\psi .
\end{equation}

Thus the $\mathrm{SU}(2)_k$ WZW model results from (\ref{eq:Z-inter}) in the massless limit ($M^2 = 0$). In the opposite extreme of sending $M^2 \to \infty$, the field $\psi(x)$ gets nailed down at the zero $\psi = \pi/2$ of $M^2 \cos^2 \psi$. In that limit, one retrieves the $\mathrm{O}(3)$ nonlinear $\sigma$-model with $\theta = \pi k$ and $\beta = 2k$.

Let us now present an abridged version of the argument \cite{AH87} that leads from $\mathrm{SM}_{\pi}$ to $\mathrm{SU}(2)_{k=1}$ WZW:

\noindent (i) The renormalization-group fixed point of $\mathrm{SM}_{\pi}$ is expected to be a conformal field theory.

\noindent (ii) The global $\mathrm{O}(3)$ symmetry of the $\sigma$-model entails three conserved Noether currents, $j_a$ ($a = 1, 2, 3$), which surely persist under renormalization. Exploiting the strong property of conformal symmetry at the RG-fixed point, one makes an argument \cite{Affleck85} that along with the $j_a$ also the Hodge-dual currents must be conserved:
    \begin{equation}\label{eq:Hodge}
    \partial_\mu j_a^\mu = 0 \; \Rightarrow \; \partial_\mu \epsilon_{\; \nu}^\mu j_a^\nu = 0 \quad (a = 1, 2, 3) ,
    \end{equation}
i.e., the conformal symmetry dictates that divergence-free currents must also be rotation-free.

\noindent (iii) The conserved currents $j_a^\mu$ and $\epsilon_{\nu}^\mu j_a^\nu$ (or rather, their holomorphic and anti-holomorphic combinations) generate an $\mathfrak{su}(2)_L \times \mathfrak{su}(2)_R$ current algebra. Given the setting with non-Abelian symmetry, this current algebra heavily restricts the set of possible RG-fixed point theories, and it is thought that such a current algebra is uniquely realized by an $\mathrm{SU}(2)$ WZW model (with some level $k$).

\noindent (iv) Owing to an anomaly of the partition function's symmetry under modular transformations, one expects \cite{Oshikawa17} that the current-algebra level can change only by steps of two ($k \to k-2$) under renormalization.

\noindent (v) Among the discrete set of $\mathrm{SU}(2)_k$ WZW models only the one with level $k = 1$ is stable with respect to generic perturbations preserving criticality.

This concludes our sketch of the argument suggesting that $\mathrm{SM}_{\pi}$ renormalizes to the $\mathrm{SU}(2)_1$ WZW model. It should be mentioned that Affleck \cite{Affleck85} has outlined a more direct derivation of $\mathrm{SU}(2)_1$ WZW for the extreme quantum case of spin $|S| = 1/2$, by starting from the 1D Hubbard model at half-filling and using non-Abelian bosonization of free fermions. In the present work, however, we consider the semiclassical limit of large $|S|$.

Now we have two critical questions about the scenario outlined above. For one, how can it happen that the mass of the field $\psi$, which is a \emph{non-Goldstone} degree of freedom (i.e.\ not low-energy protected by the global $\mathrm{O}(3)$-symmetry of the $\sigma$-model), renormalizes all the way from $M^2 = \infty$ for $\mathrm{SM}_\pi$ to $M^2 = 0$ for $\mathrm{SU}(2)_{k=1}$ WZW? Affleck and Haldane argue \cite{AH87} that the mass term $M^2 \cos^2 \psi$ is RG-irrelevant as an \emph{infinitesimal} perturbation at the RG-fixed point for $k = 1$, which is true. However, in order for that irrelevance to take effect as a property of the massless WZW model, the mass must first come down from infinity to a small neighborhood of zero. What is the agent driving such an extreme RG flow?

Our second question revolves around a number of other nonlinear $\sigma$-models at $\theta=\pi$, which are still looking for their RG-fixed points. These include the field-theory models that describe quantum phase transitions between disordered free-fermion topological insulators and superconductors in two space dimensions (there exist five of these, known as types $A$, $C$, $D$, $A$II, $D$III in the symmetry classification of \cite{HHZ05}). Some others arise from the discrete spin-chain setting by varying the symmetry group and its representation on the on-site Hilbert space. Traditionally thought to be of first order \cite{Affleck88}, some phase transitions in these generalized AF quantum spin chains might actually have a divergent correlation length \cite{PDS22}. For each of these, a scenario of Affleck-Haldane WZW-type has been ruled out (see \cite{CFT-IQHT} for an in-depth discussion of the case of the integer quantum Hall transition). So, for many if not all of these models there remains the puzzle of what happens to the field theory under renormalization to the infrared regime.

Spurred by all these questions, we do not take the main-stream scenario as a foregone conclusion but turn to an alternative description of the infrared limit of the $\mathrm{O}(3)$ model at $\theta = \pi$. Later on, we shall indicate how we envisage generalizations of that alternative description.

\section{$\mathrm{U}(1)$ boson at radius $r = 1/\sqrt{2}$}
\label{sect:U(1)}

The $\mathrm{SU}(2)_1$ WZW model makes a number of predictions about the infrared behavior of antiferromagnetic quantum spin chains with half-integer spin $|S| \gg 1$ or, equivalently, about the $\mathrm{O}(3)$ nonlinear $\sigma$-model at $\theta = \pi$. To mention one of these, the spin-spin correlation function is predicted to fall off according to the universal law
\begin{equation}\label{eq:spinspin}
    (-1)^{n-n^\prime} \langle S_n^a S_{n^\prime}^b \rangle  \sim \delta^{ab} \, |n-n^\prime|^{-1}
\end{equation}
 (independent of $|S|$) in the limit of large separation $n - n^\prime$. That algebraic decay comes about because the alternating spin $(-1)^n S_n^a$ of the quantum spin chain translates to $\mathrm{Tr}\, \sigma_a g$ as the leading operator in the continuum field theory, and the latter has scaling dimension $1/2$.

A further remark is that the RG flow of the standard scenario \cite{AH87} implies a change of target-space topology: the second homotopy group changes from $\pi_2 (\mathrm{S}^2) = \mathbb{Z}$ for the $\sigma$-model to $\pi_2(\mathrm{S}^3) = 0$ for the WZW model. The triviality of $\pi_2 (\mathrm{S}^3)$ means that the WZW model does not feature field configurations which are known as instantons and do exist as a non-perturbative feature of importance in the $\mathrm{O} (3)$ nonlinear $\sigma$-model. To rationalize the topology change, one may observe that topological excitations such as instantons tend to disorder the system, ultimately causing (exponential?) decay of correlations. Therefore, if $\mathrm{SM}_\pi$ is to renormalize to a massless limit in the infrared, then renormalization should somehow act to suppress its instantons.

Now in the community of CFT experts, it is known \cite{Ginsparg88} that the $\mathrm{SU}(2)_1$ WZW model has an equivalent representation as a free $\mathrm{U}(1)$ boson compactified at a certain radius ($r = 1/\sqrt{2}$, using string theory conventions) which makes for a hidden $\mathrm{SU}(2)$ symmetry. Taking the boson field to be an angular variable $\phi \sim \phi + 2\pi$, we write the free boson action as
\begin{equation}\label{eq:U1-S}
    S_\ast = \frac{1}{4\pi} \int_\Sigma d^2 x \, \partial^\mu \phi \, \partial_\mu \phi \,.
\end{equation}
For an alternative expression, one could standardize the prefactor in (\ref{eq:U1-S}) to the value $1/(2\pi)$ and change the scale of the angular field so that  $\phi \sim \phi + 2\pi r$ with $r = 1/\sqrt{2}$.

The model (\ref{eq:U1-S}) is equivalent to the  $\mathrm{SU}(2)_{1}$ WZW model in that it has the same conformal charge ($c = 1$) and houses the same $\mathfrak{su}(2)_L \times \mathfrak{su}(2)_R$ current algebra of level $k=1$. Indeed, splitting the angular field $\phi$ (with equation of motion $\partial_z \partial_{\bar z} \phi = 0$) into its holomorphic and anti-holomorphic parts as
\begin{equation}
    \phi(z,\bar{z}) = \frac{1}{2} \big( \phi(z) + \bar\phi (\bar{z}) \big) ,
\end{equation}
one has a triplet $\{ J^3, J^+, J^- \}$ of currents of conformal dimension $(1,0)$ which generate an $\mathfrak{su}(2)_L$ chiral current algebra of level $k=1$:
\begin{equation}
    J^3(z) = - \mathrm{i} \partial_z \phi(z) , \quad J^\pm (z) = \; :\! \mathrm{e}^{\pm \mathrm{i} \phi(z)} \! : \,.
\end{equation}
(For the $\mathfrak{su}(2)_R$ factor, simply replace $\phi(z)$ by $\bar\phi(\bar{z})$.)

Correlation functions in the two theories are also the same. In particular, the fundamental vertex field $\mathrm{e}^{\pm \mathrm{i} \phi(z,\bar{z})}$ has conformal dimension $(1/4,1/4)$, so that
\begin{equation}
    \big\langle \mathrm{e}^{\mathrm{i} \phi(z,\bar{z})} \mathrm{e}^{- \mathrm{i} \phi(w,\bar{w})} \big\rangle \sim |z-w|^{-1} ,
\end{equation}
which reproduces the scaling limit (\ref{eq:spinspin}) predicted for the spin-spin correlation function in the ground state of the antiferromagnetic chain.

So, we are faced with an instance of the same critical infrared physics being exhibited by two different models: $\mathrm{SU}(2)_{k=1}$ WZW and $\mathrm{U}(1)_{r = 1/\sqrt{2}}\,$. Of course, a major difference is that the $\mathrm{SU}(2)$ symmetry of the WZW model is not manifest in the $\mathrm{U}(1)$ boson formulation. One might therefore think that the latter is an accident which can be safely ignored. We disagree! As a matter of fact, we are going to argue that the $\mathrm{U}(1)_{r = 1/\sqrt{2}}$ description points us to a systematic phenomenon of which the present case is just one example. Let us first show how $\mathrm{SU}(2)$ can be made more explicit in the $\mathrm{U}(1)$ boson model.

\subsection{Extending the $\mathrm{U}(1)$ boson theory} \label{sect:extend}

We imagine that the angular field $\phi$ in Eq.\ (\ref{eq:U1-S}) parameterizes a circle $\mathrm{S}^1$ (the ``equator'') in the target sphere $\mathrm{S}^2$ of the $\mathrm{O}(3)$ nonlinear $\sigma$-model. For reasons to be explained presently, we remove from $\mathrm{S}^2$ two points, say the north pole $p$ and the south pole $p^\prime = - p$ (as the antipode of $p$), and we then identify the punctured sphere with the normal bundle of $\mathrm{S}^1 \subset \mathrm{S}^2$:
\begin{equation}
    \mathrm{S}^2 \setminus \{ p , -p \} \cong \mathcal{N}(\mathrm{S}^1) ,
\end{equation}
which has the topology of a cylinder, $\mathcal{N}(\mathrm{S}^1) \cong \mathrm{S}^1 \times \mathbb{R}$; in particular, $\pi_2 \big( \mathcal{N}(\mathrm{S}^1) \big) = 0$. To parameterize the cylinder axis $\mathbb{R}$, we introduce a real-valued field $b$, and we consider the extended action functional
\begin{equation}\label{eq:S-ext}
    S =\frac{\beta}{8\pi} \int_\Sigma d^2 x \, \left( \partial^\mu \phi \, \partial_\mu \phi + \nabla^\mu b \, \nabla_\mu b \right)
\end{equation}
with covariant derivative $\nabla_\mu = \partial_\mu + \mathrm{i}\, \partial_\mu \phi$. We think of this action as coming from the cylinder $\mathrm{S}^1 \times \mathbb{R}$ viewed as a Riemannian manifold with complex (!) metric
\begin{equation}\label{eq:metric}
    g = d\phi^2 + (db + \mathrm{i} b \, d\phi)^2 .
\end{equation}
Since $\mathrm{Det}(g) = 1$ in the chosen $\phi, b$ coordinate basis, the functional integration measure of the extended theory derives from the flat area two-form $d\phi \wedge db$ on $\mathrm{S}^1 \times \mathbb{R}$. We note that the factor multiplying the second summand (which has been set to unity for convenience) of the metric $g$ is arbitrary as the free field $b$ admits any change of scale $b(x) \to r \, b(x)$. As a further remark, we are not disturbed by the appearance of an imaginary term in $g$, as the topological term of the $\mathrm{O}(3)$ nonlinear $\sigma$-model as the parent theory is already imaginary. Along similar lines, we can draw inspiration from the WZW model, where the Riemannian structure of $\mathrm{SU}(2)$ is deformed by the appearance of an imaginary 2-form term introducing torsion into the geometric operation of parallel transport.

One could still worry that the functional integral of the theory (\ref{eq:S-ext}) might not exist, as the metric tensor not only fails to be real but also has one sign-indefinite diagonal component:
\begin{equation}
    \begin{pmatrix} g_{\phi \phi} &g_{\phi b} \cr g_{b \phi} &g_{bb} \end{pmatrix}
    =  \begin{pmatrix} 1 - b^2 &\mathrm{i} b \cr \mathrm{i} b &1 \end{pmatrix} .
\end{equation}
However, the functional integral does make perfect sense as long as it is understood as a sequential integral: we integrate over $b$ first, then we do the $\phi$-field integral. Indeed, carrying out the Gaussian functional integral over $b$ (after isolating the zero mode, which needs separate treatment) we obtain the reciprocal square root of a determinant:
\begin{equation}
    \begin{split}
    &\int \mathcal{D}b \; \mathrm{e}^{- (\beta / 8\pi) \int d^2x \, \nabla^\mu b \, \nabla_\mu b} \cr &\propto \mathrm{Det}^{-1/2}\big( - \mathrm{e}^{\mathrm{i} \phi} \partial^\mu \, \mathrm{e}^{- 2\mathrm{i} \phi} \partial_\mu \, \mathrm{e}^{\mathrm{i}\phi} \big),
    \end{split}
\end{equation}
which can be calculated exactly (e.g., by the method of heat-kernel regularization) and contributes to the action for the $\phi$-field the term $S_\ast$ exhibited in Eq.\ (\ref{eq:U1-S}). Thus the effective action that results from Eq.\ (\ref{eq:S-ext}) by integrating out the $b$-field is
\begin{equation}\label{eq:S-eff}
    S_{\rm eff} = \frac{\beta + 2}{8\pi} \int_\Sigma d^2 x \, \partial^\mu \phi \, \partial_\mu \phi \,.
\end{equation}
We see that the coupling to the $b$-field \emph{enhances} the stiffness of the $\phi$-field (instead of destabilizing it, as one might have feared from the metric tensor component $g_{\phi \phi} = 1 - b^2$ becoming negative for large $b$). We also see that $S_{\rm eff}$ reduces to $S_\ast$ in the limit of $\beta \to 0$.

In the next subsection, we are going to show that two-loop perturbative renormalization of the theory (\ref{eq:S-ext}) sends the coupling $\beta$ toward zero. Assuming this result to remain qualitatively correct beyond two-loop order, we arrive at the conclusion that (\ref{eq:S-ext}) is an exact reformulation of the $\mathrm{U}(1)$ boson theory (\ref{eq:U1-S}) by extension. The merit of our reformulation is that it makes the hidden $\mathrm{SU}(2)$ symmetry of (\ref{eq:U1-S}) explicit, as follows.

There  exists an infinitesimal action of
\begin{equation}
    \mathfrak{su}(2) \equiv \mathrm{Lie}\, \mathrm{SU}(2) = \mathrm{Lie}\, \mathrm{SO}(3)
\end{equation}
[actually, of the complexification $\mathfrak{sl}_2(\mathbb{C}) = \mathfrak{su}(2) \oplus \mathrm{i}\, \mathfrak{su}(2)$] on the cylinder $\mathrm{S}^1 \times \mathbb{R}$ with coordinates $\phi , b$ by three basic vector fields indexed by $\sigma_3/2$ and $\sigma_\pm = (\sigma_1 \pm \mathrm{i} \sigma_2)/2$:
\begin{equation}\label{eq:killing}
    \begin{split}
    \mathcal{L}_{\sigma_3/2} = \frac{1}{\mathrm{i}}\, \frac{\partial}{\partial\phi} \,, \quad \mathcal{L}_{\sigma_-} = \mathrm{e}^{-\mathrm{i} \phi} \, \frac{1}{\mathrm{i}}\, \frac{\partial}{\partial b} \,, \cr \mathcal{L}_{\sigma_+} = \mathrm{e}^{\mathrm{i} \phi} \left( (1-b^2) \frac{1}{\mathrm{i}} \, \frac{\partial}{\partial b} - 2 b \, \frac{\partial}{\partial\phi} \right) .
    \end{split}
\end{equation}
It is easily verified that these first-order differential operators satisfy $\mathfrak{sl}_2(\mathbb{C})$ Lie bracket relations,
\begin{equation}
    [\mathcal{L}_A , \mathcal{L}_B ] = \mathcal{L}_{[A,B]} \,,
\end{equation}
and are Killing vector fields, i.e., their vector flows preserve the metric tensor (\ref{eq:metric}) and they have zero divergence with respect to the area two-form $d\phi \wedge db$ on $\mathrm{S}^1 \times \mathbb{R}$. Thus they generate global symmetries of the action functional (\ref{eq:S-ext}) and the functional integration measure. As a result, any Ward identity due to $\mathrm{SO}(3) = \mathrm{SU}(2) / \mathbb{Z}_2$ as an infinitesimal symmetry holds in the theory (\ref{eq:S-ext}).

It should be noted, however, that the infinitesimal action of $\mathrm{SO}(3)$ does note integrate to a finite group action. The reason is that the cylinder $\mathrm{S}^1 \times \mathbb{R} = \mathcal{N}(\mathrm{S}^1 \subset \mathrm{S}^2) \cong \mathrm{S}^2 \setminus \{ p , -p \}$ is not invariant under all $\mathrm{SO}(3)$ transformations. We will return to this issue in Sect.\ \ref{sect:sum}.

\subsection{Ricci flow}\label{sect:Ricci}

{}From the literature \cite{Friedan85} we know that the renormalization group (RG) flow in two-loop approximation for a 2D nonlinear Riemannian model with coupling $\beta$ and metric tensor $g_{jl}$ is given by
\begin{equation}\label{eq:RG-1L}
    \frac{d}{d\ln a} \big( \beta \, g_{jl} \big) = - \mathrm{Ric}_{jl} -
    \frac{1}{2\beta} \mathcal{R}_{jpqr} \mathcal{R}_l^{\; pqr} + \mathcal{O}(\beta^{-2}) ,
\end{equation}
where $\mathcal{R}^i_{\; jkl}$ are the components of the Riemann curvature tensor of the target space, $\mathrm{Ric}_{jl} = \mathcal{R}^k_{\; j k l}$ is the Ricci tensor, and the RG flow parameter $a$ is the short-distance cutoff of the UV-regularized field theory. Assuming (reasonably so) that this formula from perturbation theory carries over to the case of a complex metric tensor $g_{jl}$, we proceed to apply it to the situation at hand.

To compute the curvature contractions in Eq.\ (\ref{eq:RG-1L}), we first set up the Levi-Civita covariant derivative $\nabla \equiv {}^g \nabla$ associated with the metric $g$. This is expressed as
\begin{equation}
    \nabla = d + \Gamma ,
\end{equation}
where $\Gamma$ is the connection one-form (with components that are known as Christoffel symbols). By the Koszul formula of Riemannian geometry, the expression for $\Gamma$ in our target space coordinates $\phi, b$ is
\begin{equation}
    \begin{split}
    \Gamma &= d\phi \, \big( \mathrm{i} b^2 (E_b^{\; b} - E_{\phi}^{\; \phi}) + b(1-b^2) E_b^{\; \phi} - b E_\phi^{\; b} \big) \cr
    &+ db\,  \big( b  (E_b^{\; b} - E_\phi^{\; \phi}) + \mathrm{i} b^2 E_b^{\; \phi} + \mathrm{i} E_\phi^{\; b} \big) ,
    \end{split}
\end{equation}
where $E_{\phi}^{\; \phi} = \partial_\phi \otimes d \phi$, $E_b^{\; \phi} = \partial_b \otimes d \phi$, etc., are the basic tangent space endomorphisms for our coordinate basis. Now the Riemann curvature $\mathcal{R}$ follows from the Levi-Civita covariant derivative as
\begin{equation}\label{eq:Riemann}
    \begin{split}
    &\mathcal{R} = \nabla^{\wedge 2} = d \Gamma + \Gamma \wedge \Gamma = {\textstyle{\frac{1}{2}}} \mathcal{R}_{\; jkl}^i E_i^{\; j}
    dx^k \wedge dx^l \cr &= d\phi \wedge db\, \big( \mathrm{i} b \, ( E_{\phi}^{\; \phi} - E_b^{\; b}) - (1-b^2) E_b^{\; \phi} + E_\phi^{\; b} \big) .
    \end{split}
\end{equation}
Finally, computing the terms in Eq.\ (\ref{eq:RG-1L}) we find
\begin{equation}
    \frac{1}{2} \mathcal{R}_{jpqr} \mathcal{R}_l^{\; pqr}
    = \mathrm{Ric}_{jl} = g_{jl}\,.
\end{equation}
It follows that the scalar curvature is constant and positive. Thus we conclude that our target space $(\mathrm{S}^1 \times \mathbb{R} , g)$ with complex metric $g$ behaves geometrically like a (locally) symmetric space. We shall explain in Sect.\ \ref{sect:Cauchy} why and how this striking feature comes about.

To summarize our findings, the field theory with action (\ref{eq:S-ext}) is one-parameter renormalizable up to two-loop order, and its RG flow equation (\ref{eq:RG-1L}) simplifies to
\begin{equation}\label{eq:RG-simple}
    \frac{d \beta}{d\ln a} = - 1 - \frac{1}{\beta} + \mathcal{O}(1/\beta^2) .
\end{equation}
We see that the two-loop approximation predicts an RG-beta function zero at the unphysical value of $\beta = - 1 < 0$. For reasons spelled out at the end of Sect.\ \ref{sect:Cauchy}, we expect that higher-loop corrections preserve the one-parameter renormalizability and stop the RG flow as the limit $\beta \to 0$ of stability is reached.

In summary, the theory with action (\ref{eq:S-ext}) for $\beta \to 0$ is a valid alternative to the $\mathrm{SU}(2)_1$ WZW model. Indeed, it has all the required accolades including global $\mathrm{O}(3)$ symmetry to qualify as an RG-fixed point theory for the $\mathrm{O}(3)$ nonlinear $\sigma$-model at $\theta = \pi$.

\section{Heuristic RG Scenario}\label{sect:heuRG}

Now in view of the tantalizing possibility of various generalizations to long-standing open problems, we may ask a more ambitious question: can we fortify the alternative (of Sect.\ \ref{sect:U(1)}) for the RG-fixed point of $\mathrm{SM}_\pi$ by giving a complete derivation, following the RG flow from the UV cutoff scale all the way into the IR regime? Needless to say, that looks very hard and is definitely beyond the scope of the present work. Also, to keep things in perspective, let us recall that Affleck and Haldane did not derive the $\mathrm{SU}(2)_1$ WZW model; they just made a plausible case for it as reviewed in Sect.\ \ref{sect:WZW}. With these caveats in mind, let us offer the heuristics of the present section.

\subsection{Perturbative renormalization}\label{sect:RG-pert}

For a direct treatment of the $\mathrm{O}(3)$ nonlinear $\sigma$-model at $\theta = \pi$, we need some better idea of what happens under renormalization. We begin by summarizing what is known about perturbative renormalization of the model.

(i) The $\theta$-parameter does not renormalize in perturbation theory. Indeed, the topological density $\mathcal{L}_{\rm top}$ integrates to a constant (the topological charge $q$, a.k.a.\ the number of instantons), which vanishes for all field configurations taken into account in perturbation theory.

(ii) Br\'ezin, Zinn-Justin, and Guillou \cite{BZG76} showed that the $\mathrm{O}(3)$ nonlinear $\sigma$-model (for $\theta = \pi$) is renormalizable to all orders in perturbation theory.

(iii) It should be understood that if a field-theoretical model is renormalizable to all orders in perturbation theory, it does not follow (not in any rigorous sense) that the model is actually renormalizable outside of the perturbative regime of weak coupling. In fact, perturbation theory generates but an asymptotic expansion, and unless that expansion is \emph{strongly} asymptotic, it does not determine the function which it is trying to reproduce.

(iv) For the $\mathrm{O}(N)$ nonlinear $\sigma$-model Wegner \cite{Wegner90} and Castilla \& Chakravarty \cite{CC97} computed the anomalous scaling dimensions $y_s$ of high-gradient operators of the sort $(\partial_\mu \vec{m} \cdot \partial_\nu \vec{m})^s$ by $2+\epsilon$ expansion and found the following result:
\begin{equation}\label{eq:high-grad}
    y_s = d - 2s + \frac{\epsilon s(s-1)}{N-2} + \mathcal{O}(\epsilon^2) .
\end{equation}
One notices that the scaling dimensions (\ref{eq:high-grad}) become positive for large $s \sim \epsilon^{-1}$. This led to speculations, at the time, that the nonlinear $\sigma$-model might be RG-unstable with respect to perturbations by high-gradient operators, leading to a breakdown of one-parameter scaling. Later,  Brezin and Hikami \cite{BH97} dismissed such speculations as mathematically unfounded. Nonetheless, as an afterthought to that old discussion, we find it worth remarking that results for critical exponents (of the transition in $d = 3$ to a symmetry-broken phase) from $(2+\epsilon)$-expansion of the nonlinear $\sigma$-model tend to be numerically poor in comparison with those obtained by $(4-\epsilon)$-expansion of a symmetry-related linear model of Wilson-Fisher or Ginzburg-Landau type. Therefore, and for further reasons spelled out below, another take on the situation might be that one-parameter scaling holds just fine (in agreement with numerical and experimental observations), but the 3D nonlinear $\sigma$-model is RG-unstable after all and flows to another fixed point (of Wilson-Fisher type, or of nonlinear type with a different symmetry orbit) at the phase transition.

(v) Renormalizability of the nonlinear $\sigma$-model would imply one-parameter scaling (as seen in numerics and experiments), but the converse is not true.

\subsection{Kadanoff block-spin transformation} \label{sect:RG-thoughts}

What can we say beyond perturbation theory? Let us begin with a quick word about the one-dimensional situation: nonlinear $\sigma$-models in $d=1$ are super-renormalizable by the principle of infinite divisibility of Brownian motion and its heat kernel. This simplicity is unique to one dimension and gives little (if any) insight to what happens with nonlinear $\sigma$-models under renormalization in higher dimension.

Now for the model under consideration, momentum-space RG methods are ruled out by the hard constraint $\vec{m}(x)^2 = 1$, which does not behave nicely under Fourier transform $\vec{m}(x) \mapsto \sum_x \mathrm{e}^{\mathrm{i} k x} \vec{m}(x)$. Hence we are left with real-space renormalization (of the lattice-regularized field theory) as the only viable RG scheme to use. The first idea that might come to mind is renormalization by naive decimation, i.e., by integration over some subset of the spins to produce an effective action for the remaining spins in the complementary subset. While that works fine in $d=1$, it is a bad idea in higher dimension. The reason is that, similar to what happens for renormalization by momentum-shell integration with a hard momentum cutoff, decimation will generate \emph{long-range interactions} between the remaining spins, thereby thwarting the whole renormalization approach.

So, for our nonlinear model in dimension $d \geq 2$, we are going to think about real-space renormalization by a kind of Kadanoff block-spin transformation. To set the stage, let us sketch briefly how one might implement the Kadanoff block-spin scheme in the concrete situation at hand.  (i) One discretizes the 2D continuum field theory (\ref{eq:Z-NLsM}) by placing the spins $\vec{m}(x)$ on the sites of a square lattice. The action functional then becomes a sum over interactions of nearest-neighbor pairs:
\begin{align}\label{eq:latt-act}
    &S_{\rm latt} = \frac{1}{8\pi} \sum_{x\in \mathbb{Z}^2} \sum_{\mu =1}^2 \big( \vec{m}(x + e_\mu) - \vec{m}(x)\big)^2 , \cr &\vec{m}(x) \cdot \vec{m}(x) = \beta .
\end{align}
Note that we have standardized the coupling (to $1/8\pi$) by re-interpreting the parameter $\beta$ as the radius squared of the target sphere. (We have also dropped the topological term for now with the understanding that the field is being renormalized in a sector of fixed instanton number.) (ii) One organizes the lattice sites into a super-lattice of square blocks ($b$), each containing $n^2$ sites:
\begin{equation}
    \{ \vec{m}(x + i e_1 + j e_2) \}_{1 \leq i, j \leq n} \equiv \{ \vec{m}_{b,1}, \ldots, \vec{m}_{b,n^2} \} .
\end{equation}
 (iii) One defines a suitable average over the spins in a given block $b$:
\begin{equation}
    \vec{M}_b \equiv \{ \vec{m}_{b,1} , \ldots, \vec{m}_{b,n^2} \} \mapsto \mathrm{Av}(\vec{M}_b) .
\end{equation}
(iv) One introduces collective variables $\vec{\mathbf{m}}_b$ associated with blocks by inserting
\begin{equation}\label{eq:Kadanoff}
    1 = \prod_{\rm blocks} \int d\vec{\mathbf{m}}_b \, \delta \big( \vec{\mathbf{m}}_b - \mathrm{Av}(\vec{M}_b) \big)
\end{equation}
in the functional integral. (Instead of the $\delta$-distribution one could also use a smooth bump function. The range and measure for the $\vec{\mathbf{m}}_b$ remain to be specified.) (v) Reversing the order of integration, one integrates over the variables $\vec{m}(x)$ on the original lattice to obtain a renormalized theory on the coarse-grained lattice of block variables $\vec{\mathbf{m}}_b$. (vi) One iterates the whole procedure.

\subsection{Choice of averaging map}\label{sect:arithmean}

The success of such a renormalization scheme hinges on a good choice of averaging map $\mathrm{Av}$ (and the prior question of where to concentrate the block spins $\vec{\mathbf{m}}_b$.) So, how should we choose that averaging map? Given the Riemannian structure of the target sphere $\mathrm{S}_\beta^2$ with squared radius $\beta$, one might try to place $\mathrm{Av}( \vec{M}_b )$ right on  $\mathrm{S}_\beta^2$, assign Riemann normal coordinates (w.r.t.\ to $\mathrm{Av}( \vec{M}_b)$ as the reference point) to the site spins in the given block $b$, and then determine $\mathrm{Av}(\vec{M}_b)$ by requiring that these coordinates add up to zero. While this would be a natural choice in $d=1$, we do not consider it a good choice in higher dimension, for several reasons. For one, there is again the risk of losing locality in the renormalized field theory: similar to defects, badly placed block spins will introduce an undue amount of strain into the combined system of interacting site spins and block spins, and that strain is likely to produce non-local stress among the effective system of block spins. For another reason, Riemann normal coordinates are defined only locally (they become ill-defined at the radius of injectivity of the exponential map), and this creates an issue at strong coupling where the whole target $\mathrm{S}_\beta^2$ is explored by large fluctuations of the site spins in a single block.

A better choice of the map $\mathrm{Av}$ is suggested by the (perturbatively known) outcome of a renormalization group transformation, as follows. Let us regard the sphere $\mathrm{S}^2_{\beta}$ as a submanifold of the Euclidean vector space $\mathbb{R}^3$ and take $\mathrm{Av}$ to be the arithmetic mean computed in $\mathbb{R}^3$:
\begin{equation}\label{eq:def-Av}
    \mathrm{Av}( \vec{M}_b ) = \mathrm{Av} \big( \{ \vec{m}_{b,1}, \ldots, \vec{m}_{b,n^2} \} \big) = \frac{1}{n^2} \sum_{i=1}^{n^2} \vec{m}_{b,i} \,.
\end{equation}
We then define a field-dependent scale factor $r(\vec{M}_b)$ by
\begin{equation}\label{eq:rfac}
    \mathrm{Av}( \vec{M}_b )^2 = r(\vec{M}_b) \beta .
\end{equation}
By the convexity of the 3-ball bounded by $\mathrm{S}_\beta^2$ one has $r(\vec{M}_b)  \leq 1$. Thus, the arithmetic mean (\ref{eq:def-Av}) always lies \emph{inside} the target sphere for the site spins (Fig. \ref{fig:F1}).

To calibrate our renormalization step, let us take a look at the extreme weak-coupling limit ($\beta \to \infty$). There, separated by typical distances of order unity due to the standardized weight function $\mathrm{e}^{- (\Delta \vec{m})^2 / 8\pi}$, all site spins in a given block lie effectively in a single tangent plane at a single point $p$ of the target sphere $\mathrm{S}^2_{\beta \to \infty}$ with infinite radius. Since the arithmetic average $\mathrm{Av} (\vec{M}_b)$ lies in the same tangent plane, the whole calculation admits reduction to that very plane $T_p \mathrm{S}^2_{\infty} \cong \mathbb{R}^2$. We can guess the result of the renormalization step of integrating over the site spins: by the RG-fixed point property of the 2D Gaussian free field, the renormalized action for the block spins will be given by the block analog of Eq.\ (\ref{eq:latt-act}), with exactly the same pre-factor $1/8\pi$.

As we go away from the weak-coupling limit, Friedan's formula (\ref{eq:RG-1L}) informs us that the stiffness parameter $1/8\pi$ gets reduced under renormalization, since the curvature of the target sphere $S_\beta^2$ ($\beta < \infty$) is positive. An equivalent scheme, preferred for present purposes, is to keep $1/8\pi$ fixed and encode the effect of renormalization in a shrinking squared radius $\beta$ of the target sphere. Now, even if we had no knowledge of (\ref{eq:RG-1L}), this outcome of a shrinking radius would still be expected from Eq.\ (\ref{eq:rfac}) and  $r(\vec{M}_b)  < 1$ (assuming $1/8\pi$ fixed). Indeed, a shrinking radius with reduction factor $r < 1$ is just what is predicted by the choice of arithmetic mean (\ref{eq:def-Av}) as the Kadanoff block-averaging function. It should also be clear that the reduction $\beta \to r \beta$ by arithmetic averaging simply reflects the positive target-sphere curvature, which determines the perturbative RG flow by Eq.\ (\ref{eq:RG-1L}).

\begin{figure}
    \centering
    \includegraphics[width=8cm]{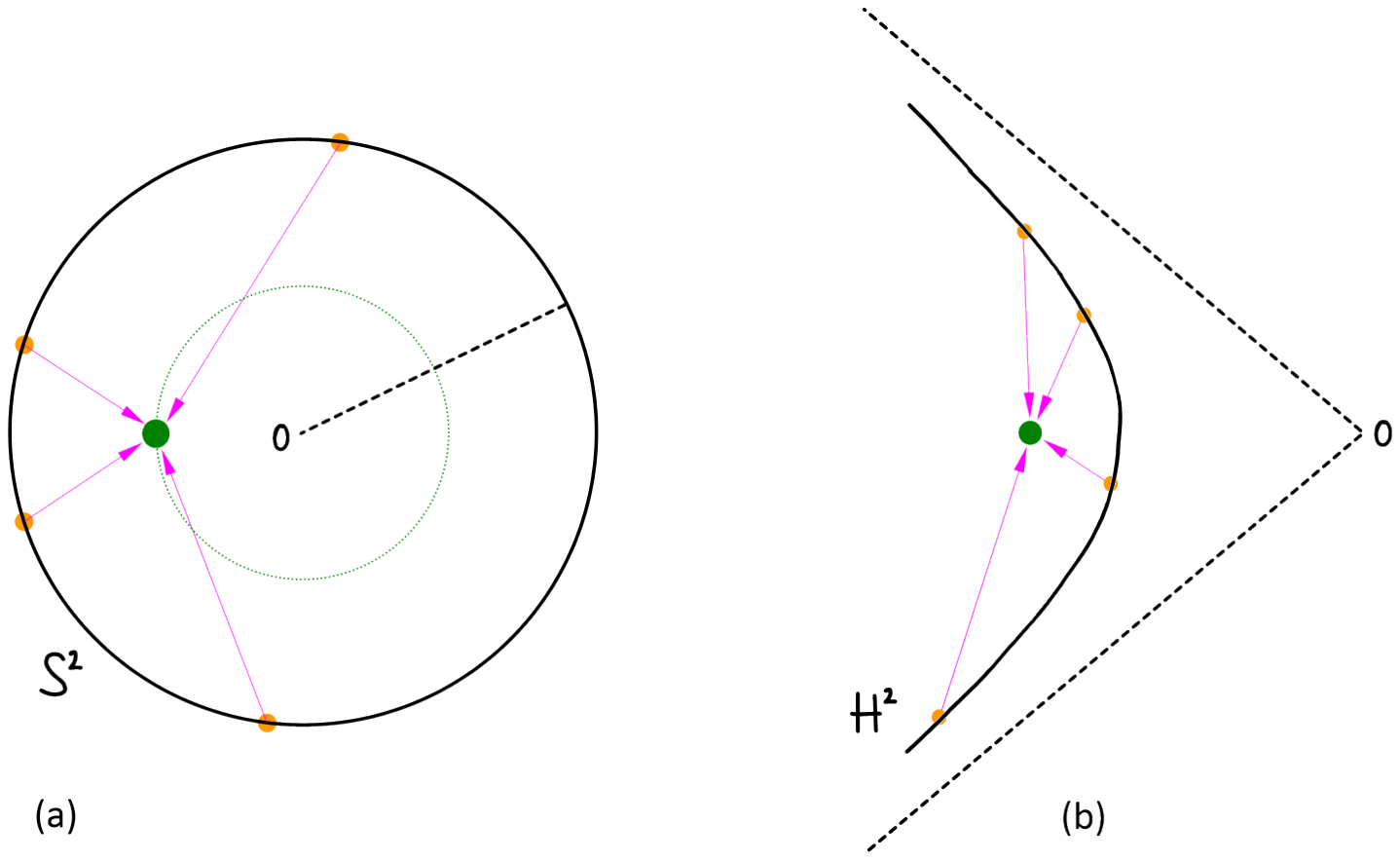}
    \caption{(a) In the case of the sphere $\mathrm{S}^2$ as the target space, the arithmetic mean (\ref{eq:def-Av}) places the Kadanoff block spin in the convex domain of the 3-ball bounded by $\mathrm{S}^2$. (b) For a hyperboloid $\mathrm{H}^2$ as target space, the same arithmetic mean still maps into a convex domain, now the 3-fold bounded by $\mathrm{H}^2$. In the former case the stiffness parameter $\beta$ (viewed as the squared target radius) decreases, while in the latter case it increases. In two dimensions, this predicts RG flow to strong resp.\ weak coupling in the two cases, in qualitative agreement with the formula (\ref{eq:RG-1L}) from perturbative RG.} \label{fig:F1}
\end{figure}

To appreciate the power of this observation, let us comment on the general situation of a nonlinear $\sigma$-model with target space $G/K$. There, to carry out the present construction, we embed the symmetric space $G/K$ as a (co-)adjoint $G$-orbit into the Lie algebra of $G$ and take the arithmetic average in the vector space $\mathrm{Lie}(G)$. As an example, consider the hyperbolic target space $\mathrm{H}^2_\beta \cong \mathrm{SO}(1,2) / \mathrm{SO}(2)$, realized as a two-dimensional subspace of the $(2+1)$-dimensional Minkowski space $\mathrm{Lie} \, \mathrm{SO}(1,2) \cong \mathbb{R}^3$ by the quadratic equation
\begin{equation}
    \begin{split} (m^0)^2 - (m^1)^2 - (m^2)^2 = \beta , \cr (m^0,m^1,m^2) \in \mathbb{R}^3 , \quad m^0 > 0 . \end{split}
\end{equation}
As before, the values of the arithmetic mean (\ref{eq:def-Av}) lie in a \emph{convex} space, now the 3-fold bounded by $\mathrm{H}^2_\beta$. In the present instance, this means that the scale factor $r(\vec{M}_b)$ always \emph{exceeds} unity (cf.\ Fig.\ \ref{fig:F1}). Thus the squared radius $\beta$ (or ``invariant mass'', using the language of special relativity) is now predicted to grow under renormalization. This is how it should be, given the formula (\ref{eq:RG-1L}) and the fact that $\mathrm{H}^2_\beta$ has constant \emph{negative} curvature.

In summary, the arithmetic mean (\ref{eq:def-Av}) provides us with a geometric and very direct intuition for how renormalization affects a 2D nonlinear $\sigma$-model: for a target space of positive (resp.\ negative) curvature, it correctly predicts a shrinking (resp.\ expanding) target radius, hence RG flow to strong (resp.\ weak) coupling. We take this as convincing evidence that the arithmetic mean  (\ref{eq:def-Av}) is a good averaging map to use in the Kadanoff block-spin transformation given by Eq.\ (\ref{eq:Kadanoff}).

\subsection{What happens at strong coupling?}\label{sect:punch}

With that understanding in hand, we can finally deliver the punch line of the current section. In the weak-coupling regime, we expect that one can renormalize the lattice nonlinear $\sigma$-model (\ref{eq:latt-act}) to another such model; to that end, one replaces the $\delta$-distribution in (\ref{eq:Kadanoff}) by a smooth bump distribution and constrains the block spins $\vec{\mathbf{m}}_b$ to vary in a target sphere with \emph{definite and optimized} \cite{DST24} squared radius $\beta^\prime < \beta$. Not so for strong coupling! Given the present picture, it looks quite unlikely that one can avoid generating long-range renormalized interactions with the stipulation that block spins are constrained to lie in a target sphere of \emph{fixed} radius. Rather, the radius should be allowed to fluctuate and become a dynamical variable, thereby increasing the target-space dimension from two to three -- in line with the Affleck-Haldane proposal of RG flow to the $\mathrm{SU}(2)_1$ WZW model with a three-dimensional target space.

The key message then is this. Even though the $\mathrm{O}(3)$ nonlinear $\sigma$-model is renormalizable by perturbation expansion about the Gaussian free field limit at weak coupling, there exists no proof of renormalizability in the strong-coupling regime; moreover, in view of the picture presented here, it seems highly unlikely that such will ever come about. Rather, one should accept that in the course of renormalization to strong coupling the radius of the target sphere becomes a dynamical field. (As a corollary, one should not be surprised if the model ultimately undergoes target-space reconstruction in the final stretch of RG flow to the infrared limit.)

To add further substance to our key message, let us mention the possibility of implementing the nonlinear constraint $\vec{m}(x)^2 =1$ by integration over a Lagrange multiplier field $x \mapsto \vec{\lambda}(x)$. The resulting linearized theory can be analyzed in the momentum representation by functional RG [Flore]. By doing so, one still finds that $\vec{\lambda}$ becomes dynamical, in keeping with the picture above ($2 \to 3$ d.o.f.). An alternative pursued in \cite{HSY24} is to drop the nonlinear constraint $\vec{m}(x)^2 = 1$ and, instead, let the target sphere radius $\vert \vec{m}(x) \vert$ fluctuate according to a low-order polynomial $U$ added as $\int d^2x \, U(|\vec{m}(x)|^2)$ to the bare Lagrangian. Such an approach is in full agreement with the present scenario. (We should caution, however, that the final conclusions of \cite{HSY24} are not the same as ours.)

To conclude this subsection, let us qualify the above by putting it in the grand perspective. Recall that any RG-generated spacetime fluctuations of the target-sphere radius are fluctuations of a non-Goldstone (or \emph{massive}) field not related to the global symmetries of the $\mathrm{O}(3)$ nonlinear $\sigma$-model. Therefore, we expect these to be a \emph{transient} (albeit very important) phenomenon of non-perturbative renormalization. In other words, we expect the field degrees of freedom to reconcentrate on a \emph{single} orbit of the $\mathrm{O}(3)$-symmetry group as the RG flow arrives in the far infrared; there, we are going to see no more than massless field degrees of freedom -- massless due to the continuous $\mathrm{O} (3)$-symmetry of the model.

\subsection{What to do about the $\theta$-term?}\label{sect:fate-theta}

The discussion so far ignores the $\theta$-term of the $\sigma$-model.
To incorporate it into our local RG picture, we first need to clarify the interpretation of $\theta = \pi$ (mod $2\pi$). Indeed, the field-theory models at $\theta$ and $\theta + 2\pi$ are equivalent only on a spacetime \emph{without boundary}; in the presence of a boundary, the physics is verifiably different: the model for $\theta + 2\pi$ features an additional edge mode as compared with that for $\theta$. Now, real-space renormalization is a locality-preserving procedure that processes block-type (meaning local) information. How, then, should we go about renormalizing the present theory with coupling-constant anomaly $\theta = \pi + 2\pi \mathbb{Z}$?

Recall from Sect.\ \ref{sect:H-AF} that we are studying $\mathrm{SM}_\pi$ primarily as an effective field theory for antiferromagnetic quantum spin chains in the semiclassical limit of large spin $|S|$, where the bare $\theta$-coupling is
\begin{equation*}
    \theta = 2\pi |S| .
\end{equation*}
To appreciate the large value of $\theta \sim |S|$, notice that the $\theta$-term in (\ref{eq:Z-NLsM}) arises from the path-integral method with spin-coherent states via the \emph{overlap} between such states. Relatedly, one may think of $(\theta / 2\pi) \int_\Sigma d^2x\, \mathcal{L}_{\rm top}$ as the Berry phase incurred in the course of adiabatic transport of a quantum spin $|S|$ around a closed spacetime loop bounding $\Sigma$; that Berry phase grows \emph{linearly} with $|S|$.

Having argued that the target-sphere radius turns dynamical under renormalization into the strong-coupling regime, we now face the question as to how the $\theta$-term extends away from fixed radius. Consider the extension
\begin{align}\label{eq:ext-theta}
    &\frac{\mathrm{i} \theta}{2\pi} \int_\Sigma d^2x\, \mathcal{L}_{\rm top} \to \cr &\frac{\mathrm{i}}{4\pi} \int_\Sigma d^2x \, f(|\vec{m}|^2) \, \epsilon^{\mu\nu}\, \vec{m} \cdot (\partial_\mu \vec{m} \times \partial_\nu \vec{m})
\end{align}
where $\vec{m}(x) \in \mathbb{R}^3$ and $f$ is some profile function.
To reproduce the proper semiclassical behavior for large $|\vec{m}| \sim |S|$, we require the asymptotic limit $f(|\vec{m}|^2) \to |\vec{m}|^{ -2}$ for large $|\vec{m}|$. If we also insist on an analytic dependence near $\vec{m} = 0$, then a natural choice is
\begin{equation}\label{eq:profile}
    f(|\vec{m}|^2) = (\alpha + |\vec{m}|^2)^{-1}
\end{equation}
with some positive number $\alpha$ of order unity. [Incidentally, one may observe that the Wess-Zumino-Witten term given by (\ref{eq:dynamo}) depends on the radial field $\psi$ linearly near $\psi = \pi/2$ and as a cubic power near $\psi = 0$, similar to what we have in Eqs.\ (\ref{eq:ext-theta}, \ref{eq:profile}). We do not know whether to attribute a deeper significance to that coincidence.]

As a side remark, the functional RG approach of \cite{HSY24} takes $f \equiv 1$; we do not consider that an appropriate choice of extension.

\subsection{Target-space surgery}\label{sect:surgery}

After a long excursion into non-perturbative renormalization and what to expect from it, we return to the stage of Sect.\ \ref{sect:U(1)} to tie down some loose ends. Recall that the phenomenological target space of the infrared (IR) theory (\ref{eq:S-ext}) is a cylinder $\mathrm{S}^1 \times \mathbb{R}$, whereas the target space of the UV theory is a sphere $\mathrm{S}^2$. To give further justification to the proposed IR theory, we need to come up with a scenario explaining the change of topology ($\mathrm{S}^2 \to \mathrm{S}^1 \times \mathbb{R}$). That scenario has to revolve around the $\theta$-term.

Without the input from non-perturbative renormalization, we might argue as follows. We observe that the UV model (\ref{eq:Z-NLsM}) has instanton configurations \cite{FFS79, BL79}, where the field $\vec{m}$ wraps around the target sphere. One speaks of an ``instanton gas'' of lumps of topological charge density. (The gas is actually a plasma of localized excitations carrying charges of both signs.) While such configurations are invisible to perturbative renormalization where one integrates out the short-wavelength modes, they do have the non-perturbative effect of scrambling the field over large scales. For $\theta \not= \pi$ the scrambling effect presumably helps with the generation of a mass gap, making all correlations decay exponentially.

For $\theta \not = 0$ (mod $2\pi$) the sum over instantons comes with a phase factor $\mathrm{e}^{\mathrm{i} \theta q}$, where $q \in \mathbb{Z}$ is the total topological charge; cf.\ Eq.\ (\ref{eq:topo-q}). The presence of the phase factor causes cancelations in the sum. These cancelations become maximal for $\theta = \pi$, where $\mathrm{e}^{ \mathrm{i} \theta q} = (-1)^q$. We are thus led to surmise that the parameter value $\theta = \pi$ acts to suppress the sectors with $q \not= 0$; i.e., instantons do exist on short scales, but on approaching the infrared limit the instanton plasma (for $\theta = \pi$) acquires local neutrality.

To corroborate that surmise, let us now inject our thoughts from the current section. We went to some length to argue that non-perturbative renormalization at strong coupling promotes the target-sphere radius transiently to a dynamical field. Building on that strong-coupling scenario, we expect that the $\theta$-parameter exhibits spacetime fluctuations that increase with decreasing stiffness $\beta$. These RG-induced fluctuations of $\theta$ enter the weight of the functional integral as a multiplicative phase factor
\begin{equation}\label{eq:mult-phas}
    \exp \left( \frac{\mathrm{i}}{2\pi} \int d^2x \, \theta(x) \, \mathcal{L}_{\rm top}(x) \right) .
\end{equation}
By the Fourier principle that broad distribution of a variable forces narrow distribution of the conjugate variable, the increasing fluctuations of $\theta(x)$ will suppress the topological density $\mathcal{L}_{\rm top}(x)$. That is our main hypothesis.

By inspecting the expression for $\mathcal{L}_{\rm top}$, say in spherical polar coordinates with a polar angle $-\pi/2 \leq \xi \leq +\pi/2$ and an azimuthal angle $0 \leq \phi \leq 2\pi$,
\begin{equation}
    \mathcal{L}_{\rm top} = {\textstyle{\frac{1}{2}}} \, \epsilon^{\mu\nu} \cos \xi \, \partial_\mu \xi \, \partial_\nu \phi ,
\end{equation}
we see that the topological density stays small if at least one of $\xi$ or $\phi$ exhibits small variation. Now for a fixed field configuration we can always change to adapted coordinates to arrange for $\xi$ to be the variable that exhibits restricted variation. The emerging picture then is that (in a given field configuration of statistical relevance) the field explores a tubular neighborhood of some equator $\xi = 0$ while staying away from two polar cap regions $\xi \geq \pi/2 - \delta$ and $\xi \leq -\pi/2 + \delta$. Assuming this picture to be qualitatively correct, we may excise from the target sphere $\mathrm{S}^2$ the two dark polar caps. Actually, for our purposes it will be sufficient to remove two points (the ``north pole'' $p$ at $\xi = \pi/2$ and the ``south pole'' $-p$ at $\xi = -\pi/2$), thus replacing  $\mathrm{S}^2$ by $\mathrm{S}^2 \setminus \{ p , -p \}$. This step of excision is a necessary prerequisite for the remaining part of our treatment: we shall now Cauchy-deform the $\mathrm{O}(3)$ nonlinear $\sigma$-model with modified target space $\mathrm{S}^2 \setminus \{ p , -p \}$ to the field theory (\ref{eq:S-ext}) with cylindrical target space.

\begin{figure}
    \centering
    \includegraphics[width=8cm]{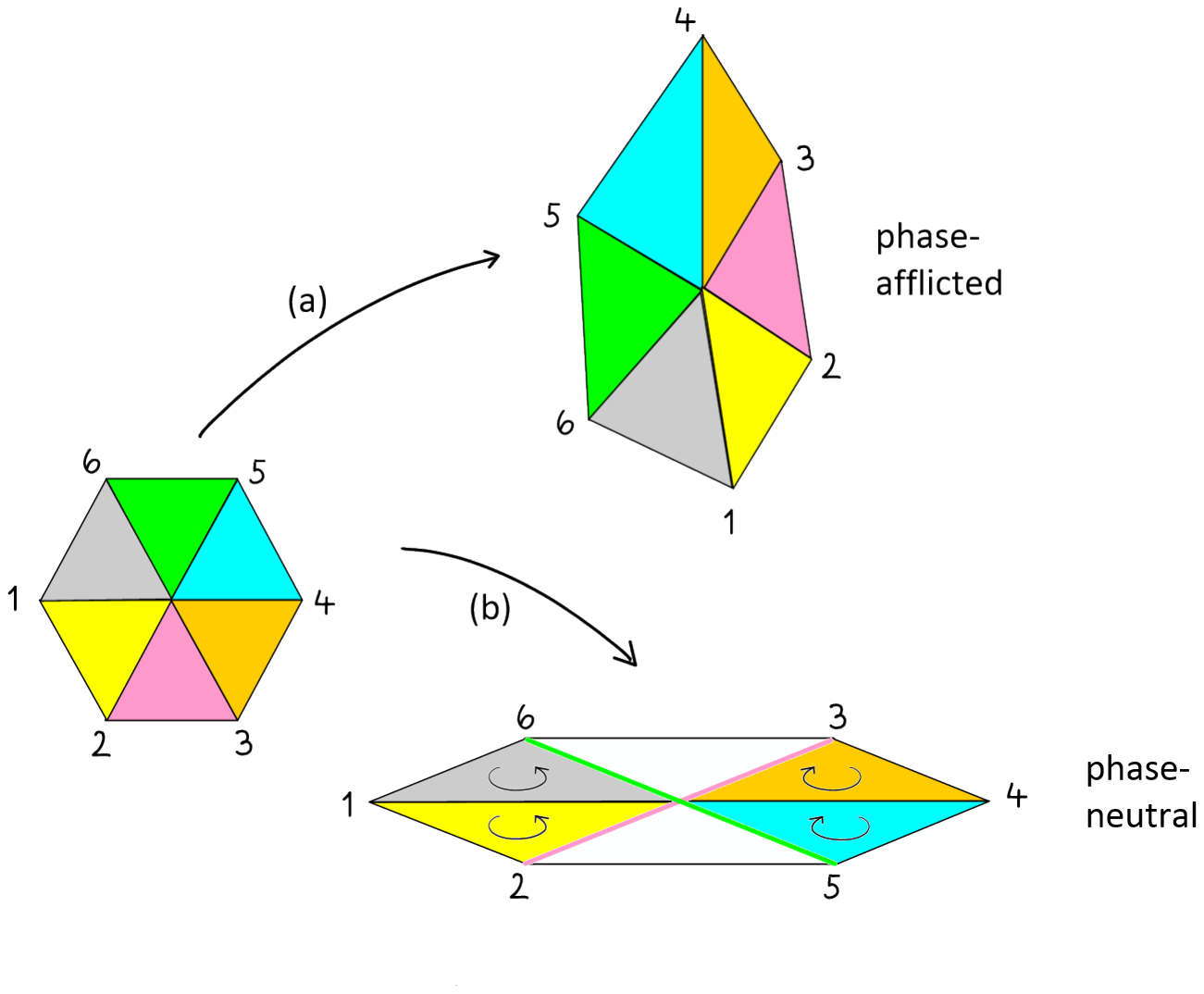}
    \caption{(a) Orientation-preserving maps from $\Sigma$ to $\mathrm{S}^2$ accumulate a large phase $(\theta/2\pi) \int_\Sigma d^2x \, \mathcal{L}_{\rm top}$, which is expected to cause a large amount of cancelation by destructive interference. (b) In contrast, field configurations that undulate in some equatorial strip of the target sphere by multiple orientation reversals, typically are close to being phase-neutral. Such configurations should dominate the infrared.}
    \label{fig:F2}
\end{figure}

\section{Cauchy deformation}\label{sect:cauchy}

A basic and well known result of complex analysis on $\mathbb{C}$ is the Cauchy Integral Theorem. One formulation of it is that line integrals admit contour deformation, i.e.
\begin{equation}\label{eq:cauchy}
    \int_\gamma f(z) \, dz = \int_{\gamma^\prime} f(z) \, dz ,
\end{equation}
if the two curves differ only by a boundary, $\gamma^\prime = \gamma + \partial D$ (one also says that $\gamma^\prime$ is homologous to $\gamma$), and $f$ is holomorphic on the connecting domain $D$. The equality (\ref{eq:cauchy}) carries over to the setting of $\mathbb{C}^n$ and, more generally, any complex manifold. Here we adapt this ``Cauchy principle'' to our field-theory problem.

\subsection{Cauchy principle (field theory version)}\label{sect:cauchy-FT}

To motivate the field-theory generalization, let us consider some field theory with target space $\mathcal{X}$ which is a real submanifold of real dimension $n$ of a complex $n$-dimensional manifold $\mathcal{C}$ with local coordinates $z^1 , \ldots, z^n$. Let $\mathcal{C}$ be equipped with a section of its canonical bundle, i.e.\ a differential form
\begin{equation}
    \omega^{(\mathcal{C})} = dz^1 \wedge \cdots \wedge dz^n
\end{equation}
of degree $n$. Assume that $\mathcal{X} \subset \mathcal{C}$ is orientable with volume form $\omega^{(\mathcal{X})}$ given by the restriction of $\omega^{(\mathcal{C})}$.

Now imagine that the field theory is UV-regularized by placing it on a lattice, say $\mathbb{Z}^d$, or even better, a finite part $\Lambda \subset \mathbb{Z}^d$ thereof. The field then becomes a map $\Phi :\; \Lambda \to \mathcal{X}$, or equivalently, a point of the product space $\mathcal{X}^{|\Lambda|}$ where $|\Lambda|$ is the cardinality of the discrete set $\Lambda$. By taking products, the lattice field space $\mathcal{X}^{ |\Lambda|}$ is embedded into $\mathcal{C}^{ |\Lambda|}$ as a real subspace.

We now assume that the field-theory integrand $\mathrm{e}^{-S}$ extends holomorphically to an open neighborhood of $\mathcal{X}^{|\Lambda|}$ inside $\mathcal{C}^{|\Lambda|}$. By the Cauchy principle, the functional integral then remains unchanged if we deform $\mathcal{X}$ to a homologous domain $\widehat{\mathcal{X}} \subset \mathcal{C}$ (hence $\mathcal{X}^{|\Lambda|}$ to $\widehat{\mathcal{X}}^{|\Lambda|}$) inside the domain of holomorphicity:
\begin{equation}
    \int_{\mathcal{X}^{|\Lambda|}} \mathrm{e}^{-S} \prod_{x \in \Lambda} \omega_x^{(\mathcal{C})} =
    \int_{\widehat{\mathcal{X}}^{|\Lambda|}} \mathrm{e}^{-S} \prod_{x \in \Lambda} \omega_x^{(\mathcal{C})} ,
\end{equation}
while keeping the integrand $\mathrm{e}^{-S} \prod \omega_x^{( \mathcal{C})}$ the same.

It stands to reason that the stated Cauchy principle continues to hold when the lattice regularization is replaced by another scheme of UV-regularization. In the sequel, we shall apply it to the continuum field theory (\ref{eq:Z-NLsM}, \ref{eq:meas}) with unspecified UV-regularization.

\subsection{Cauchy deformation of $\mathrm{S}^2 \setminus \{ p , -p \}$}\label{sect:Cauchy}

We now pause for a moment to say where we are and what we are  going to do. Recall that our renormalization group scenario is multi-stage: starting from perturbative RG at weak coupling, we pass through a strong-coupling regime with dynamical but transient target-radius field, and we finally arrive at the infrared limit with only massless degrees of freedom. In the present subsection, we are going to jump over the middle stage, for which we have nothing quantitative to offer beyond informed guess work. More precisely, we will use only one piece of information from the middle stage: the target-space surgery $\mathrm{S}^2 \to \mathrm{S}^2 \setminus \{ p , -p \}$. Thus we now consider the model (\ref{eq:Z-NLsM}) with $\vec{m}(x)^2 = 1$ but modified target space $\mathrm{S}^2 \setminus \{ p , -p \}$ to show that it deforms \emph{exactly} to the field theory (\ref{eq:S-ext}).

Continuing our notation from the previous subsection, we make the identification
\begin{equation}
    \mathcal{X} \equiv  \mathrm{S}^2 \setminus \{ p , -p \}
\end{equation}
and view the punctured two-sphere $\mathcal{X}$ as a real subspace of dimension 2 of the complex 2-dimensional quadric
\begin{equation}
    \mathcal{C} = \{ (m^1 , m^2 , m^3) \in \mathbb{C}^3 \mid \sum\nolimits_a (m^a)^2 = 1 \} .
\end{equation}
Introducing the combinations $m^\pm := m^1 \mp \mathrm{i} m^2$ and taking
\begin{equation}\label{eq:m3}
    m^3 = \pm \, \sqrt{1 - m^+ m^-}
\end{equation}
as dependent, we will regard the complex variables $m^\pm$ as local complex coordinates for $\mathcal{C}$ [in either one of the two coordinate charts with fixed sign in Eq.\ (\ref{eq:m3})]. The real subspace $\mathcal{X} \subset \mathcal{C}$ is singled out by demanding
\begin{equation}
    m^- = \overline{m^+}
\end{equation}
and excluding $m^+ = 0 = m^-$. The usual solid-angle two-form on $\mathcal{X}$ is given by restriction of
\begin{equation}
    \omega^{(\mathcal{C})} = \frac{dm^1 \wedge dm^2}{m^3} \,.
\end{equation}

It is clear that the functional integrand from (\ref{eq:Z-NLsM}, \ref{eq:meas}) is a holomorphic functional of the field $\vec{m}$ with complexified target space $\mathcal{C}$. Thus the conditions for the applicability of the Cauchy principle are satisfied, and we may deform the punctured two-sphere $\mathcal{X} = \mathrm{S}^2 \setminus \{ p , -p \}$ to a homologous target space inside the quadric $\mathcal{C}$. For notational convenience, we now assemble the components of $\vec{m}$ into the matrix $Q = \mathrm{i} \sigma_a m^a$. Then, to fix our setting, we identify the excised points $\pm p$ with $Q = \pm \mathrm{i} \sigma_3$.

The deformation is implemented by a family of differentiable maps $f_\tau : \; \mathcal{X} \to \mathcal{C}$ with real deformation parameter $\tau \geq 0$. Rather than defining $f_\tau$ directly, we find it technically convenient to specify how the local coordinates $m^\pm$ transform under inverse pullback by $f_\tau$:
\begin{align}\label{eq:tau-def}
    \begin{split}
    (f_\tau^{-1})^\ast m^+ = m^+ \cosh\tau - \sqrt{ \frac{m^+}{m^-} } \sinh\tau , \cr
    (f_\tau^{-1})^\ast m^- = m^- \cosh\tau + \sqrt{ \frac{m^-}{m^+} } \sinh\tau .
    \end{split}
\end{align}
We see that the expressions on the right-hand side are locally holomorphic in the complex variables $m^\pm$. We also note that the square root becomes single-valued upon restriction to $\mathcal{X}$, where $m^- = \overline{m^+}$. Furthermore, from the requirement that $f_\tau$ maps $\mathcal{X}$ into $\mathcal{C}$ (i.e.\ preserves the defining constraint $\vec{m}^2 = 1$), we deduce that
\begin{equation}
    ( f_\tau^{-1})^\ast m^3 = m^3 \cosh\tau .
\end{equation}

Of course, $f_\tau$ is ill-defined on the complex lines $m^+ = 0$ and $m^- = 0$. These intersect $\mathrm{S}^2 \subset \mathcal{C}$ in the two points $\{ p, - p \}$, which is why it was necessary to replace $\mathrm{S}^2$ by the punctured space $\mathcal{X} = \mathrm{S}^2 \setminus \{ p , -p \}$. Altogether, we conclude that we have a good one-parameter family of maps $f_\tau :\; \mathcal{X} \to \mathcal{C}$, giving us a family of deformed spaces $f_\tau(\mathcal{X})$ all homologous to $f_0(\mathcal{X}) = \mathcal{X}$.

We now proceed to carry out the Cauchy deformation by $f_\tau$ (ultimately sending $\tau \to \infty$). In order to avoid the unnecessary coordinate singularity at the equator $m^+ m^- = 1$ where the square root $m^3 = \pm \sqrt{1 - m^+ m^-}$ changes sign, we switch to spherical polar coordinates for $\mathcal{X}$ by setting
\begin{equation}
    m^\pm \big\vert_{\mathcal{X}} = \mathrm{e}^{\mp \mathrm{i} \phi} \cos \xi
\end{equation}
with polar angle $\xi$ and azimuthal angle $\phi$ as in Sect.\ \ref{sect:surgery} above. (The values $\xi = \pm \pi/2$ of the former are excluded by excision of the poles $\{ p , -p \}$). By the transformation (\ref{eq:tau-def}), the basic functions $m^a$ (assembled into $Q = \mathrm{i} \sigma_a m^a$) then pull back to
\begin{align}
    Q_\tau &\equiv (f_\tau^{-1})^\ast Q = \mathrm{i} \sigma_3 \sin \xi \, \cosh\tau \cr
    &+ \mathrm{i} \sigma_+ \mathrm{e}^{-\mathrm{i} \phi} (\cos \xi \, \cosh\tau + \sinh\tau) \\
    &+ \mathrm{i} \sigma_-  \mathrm{e}^{+\mathrm{i} \phi} (\cos \xi \, \cosh\tau - \sinh\tau) , \nonumber
\end{align}
and the Fubini-Study metric of the two-sphere $\mathrm{S}^2$,
\begin{equation}
    g = - \frac{1}{2} \mathrm{Tr} \, dQ^2 = d\phi^2 \cos^2 \xi + d\xi^2 ,
\end{equation}
$\tau$-deforms to ${}^\tau \! g \equiv - \frac{1}{2} \mathrm{Tr} \, d Q_\tau^2$ as
\begin{align}\label{eq:tau-metric}
    {}^\tau \! g &= d\phi^2 (1 - \sin^2 \xi \, \cosh^2 \tau) \cr
    &+ d\xi^2 \cosh^2 \tau + \mathrm{i} \sin \xi\, d\xi \, d\phi \, \sinh 2\tau .
\end{align}
Let us point out that ${}^\tau \! g$ (with its imaginary off-diagonal component) has eigenvalues that are real and positive. Indeed,
\begin{align*}
    &\mathrm{Det}\, ({}^\tau \! g) = \cos^2 \xi \, \cosh^2 \tau > 0 , \cr
    &\mathrm{Tr}\, ({}^\tau \! g) = 1 + \cos^2 \xi \, \cosh^2 \tau > 0 .
\end{align*}

We finally introduce the new variable $b := \sin \xi \, \cosh\tau$ and take the limit $\tau \to \infty$. The range of $b$ then expands to the full real axis, and the metric (\ref{eq:tau-metric}) converges pointwise (not uniformly) to the metric (\ref{eq:metric}). Dropping the $\tau \to \infty$ remnant of the $\theta$-term, which is zero due to $\pi_2(\mathcal{X}) = 0$, we conclude that the theory with action (\ref{eq:S-ext}) is equivalent to the original theory, Eq.\ (\ref{eq:Z-NLsM}). Note that we have reached that conclusion by making one (and only one!) assumption: instantons are suppressed and we may change the target-space topology from $\mathrm{S}^2$ to $\mathcal{X}$.

Moreover, we now see why the target-space geometry (\ref{eq:metric}) has constant scalar curvature and thus behaves like a locally symmetric space: being a Cauchy deformation of $\mathcal{X} = \mathrm{S}^2 \setminus \{ p , -p \}$, the target space $(\mathrm{S}^1 \times \mathbb{R},g)$ receives its geometric structure by holomorphic continuation from the Riemannian geometry of the symmetric space $\mathrm{S}^2$.

In Sect.\ \ref{sect:Ricci} we showed that Ricci flow lowers the coupling $\beta$ of the $\tau \to \infty$ deformed theory. Thus we know that the RG beta function is negative for $\beta$ large. Now in view of the hidden symmetric-space geometry and the resulting simplification of the RG flow [cf.\ Eq.\ (\ref{eq:RG-simple})] there is no reason for the RG beta function to change sign when $\beta$ becomes small. (Indeed, in that case there would be a multiplicity of RG-fixed points begging for a physical explanation that does not exist.) Thus by monotonicity, we expect the coupling to renormalize all the way down to $\beta = 0+$, where the RG flow definitely must stop to avoid running into the instability of the theory for $\beta < 0$. Our ``derivation'' of the $\mathrm{U}(1)_{r = 1/\sqrt{2}}$ boson (as an RG-fixed point theory for critical anti-ferromagnetic quantum spin chains) is then completed by recalling that integration over $b$ yields the effective action (\ref{eq:S-eff}) and hence, for $\beta \to 0$, the fixed-point action (\ref{eq:U1-S}).

It should be emphasized again that the present picture is heuristic (and, sadly, any rigorous treatment will have to involve the full apparatus of non-perturbative renormalization theory). Its merit is that it allows us to guess the solution of several non-trivial problems that have so far defied solution.

\section{Summary and Outlook}\label{sect:sum}

Since our line of reasoning has been presented in pedagogical bits and pieces distributed over a number of subsections, we shall now give a concise summary of our main thread of thought. The starting premise is that the $\mathrm{O}(3)$ nonlinear $\sigma$-model with semiclassical parameter values $\theta = 2\pi |S| = \beta$ ($S$ half-integer and large) is a critical theory. To begin the search for its non-trivial renormalization-group fixed point, we note that weak-coupling perturbation theory takes $\beta$ to smaller values, eventually sending the field theory into a strong-coupling regime where we lose analytical control.

What happens as we enter that analytically inaccessible regime? Here we take inspiration from real-space renormalization by Kadanoff block-spin transformations and also from a very plausible feature of the WZW scenario: no longer constant in space, the couplings $\beta$ and $\theta$ start fluctuating as functions of a dynamical field, as in Eq.\ (\ref{eq:dynamo}). The rationale here is that large spatial variations of the nonlinear $\sigma$-model field in the strong-coupling regime cannot be accounted for by RG transformations onto a single orbit of the $\mathrm{O}(3)$-symmetry group action; rather, the coarse-grained field wants to spread over a higher-dimensional target space foliated by such orbits. (This effect is well known in space dimension $d=3$ and higher, where one describes RG-fixed points \emph{not} as nonlinear $\sigma$-models but as fixed points of multi-orbit or Wilson-Fisher type \cite{JLM17}.)

The RG-induced fluctuations of $\theta$ enter the weight of the functional integral as the multiplicative phase factor (\ref{eq:mult-phas}), which tends to reduce the statistical weight by destructive interference. Our main conjecture then is that the dynamical fluctuations of $\theta$ act to suppress the topological charge density $\mathcal{L}_{\rm top}$, so that the instanton gas is ultimately driven to local charge neutrality in the infrared limit. If so, the topology of (a single $\mathrm{O}(3)$-orbit in) the target space gets effectively altered to that of an equatorial strip or cylinder in $\mathrm{S}^2$:
\begin{equation*}
    \mathrm{S}^2 \rightarrow \mathcal{N}(\mathrm{S}^1 \subset \mathrm{S}^2) \cong \mathrm{S}^1 \times \mathbb{R} .
\end{equation*}
How can we get from here to an RG-fixed point?

In order for the renormalization-group flow to stop, the field theory must reach a point of ``geometrostasis'' \cite{BCZ85}, i.e., the target-space geometry must flow to an RG-stationary state where the (generalized) target-space curvature vanishes. One can imagine two possibilities. For one, you may embrace the WZW scenario (with its extended target space $\mathrm{S}^3$ including the dynamical field $\psi$) to its full extent, hoping that RG transformations will somehow induce the WZW-anomalous term that is needed to offset (by torsion) the Riemannian curvature of $\mathrm{S}^3$. What remains unexplained in that scenario is how the orbit-fluctuation field $\psi$, which is non-Goldstone (i.e.\ not protected by any continuous symmetry of the $\mathrm{O}(3)$-model), escapes the fate of infrared death due to the persistence of its mass terms. We are thus motivated to advocate another possibility: the appearance of $\psi$ is but a transient effect of renormalization at strong coupling --- $\psi$ is actually massive and gets frozen out in the final stages of the RG flow into the infrared fixed point.

In the latter scenario, geometrostasis is achieved by evolution of the target-space geometry to the parallelizable space $\mathrm{S}^1 \times \mathbb{R}$ of a cylinder with the complex (!) metric tensor $g$ of Eq.\ (\ref{eq:metric}). To reach an RG-fixed point in this scenario, we take the action functional of Eq.\ (\ref{eq:S-ext}) and send the coupling $\beta \to 0+$. A painless ``derivation'' (skipping all the hard work of renormalization) of the expression (\ref{eq:S-ext}) is by Cauchy deformation of $\mathrm{S}^2 \setminus \{ p , -p \} \equiv \mathcal{X}$ to $f_{\tau\to\infty}(\mathcal{X}) \cong \mathrm{S}^1 \times \mathbb{R}$, as demonstrated in Sect.\ \ref{sect:Cauchy}.

We still need to address a final point that was left out in the discussion of Sect.\ \ref{sect:extend}: the infinitesimal action of $\mathrm{SO}(3)$ by the Killing vector fields (\ref{eq:killing}) does not integrate to a finite group action, as the excision points $\{ p , -p \}$ do not remain fixed under a general $\mathrm{SO}(3)$ transformation. Now by the Mermin-Wagner-Coleman principle, continuous compact symmetries such as the $\mathrm{SO}(3)$-symmetry of $\mathrm{SM}_\pi$ cannot be spontaneously broken in two dimensions. What should we then make of the apparent violation of global symmetry under the connected Lie group $\mathrm{SO}(3)$?

Our answer is that the global symmetry action by $\mathrm{SO}(3)$ comes with a zero mode, and we have not yet taken care of it. At the infrared RG fixed point, this can be done as follows. Given any reference point in the target sphere, $p \in \mathrm{S}^2$, we have the $\tau$-deformed target cylinder
\begin{equation}
    C_p := f_{\tau \to \infty}\big(\mathrm{S}^2 \setminus \{ p , -p \} \big) \cong \mathrm{S}^1 \times \mathbb{R} ,
\end{equation}
which we equip with (local) coordinate functions
\begin{equation}
    \phi : \; C_p \to \mathrm{R}/ 2\pi\mathbb{Z} , \quad b :\; C_p \to \mathbb{R} .
\end{equation}
Composing these with the field map from spacetime $\Sigma$ into $C_p$, we define the partition function
\begin{equation}\label{eq:Z-Cp}
    \begin{split}
    &Z(C_p) = \int \mathcal{D} \phi \int \mathcal{D}b \; \mathrm{e}^{- \int_\Sigma d^2x \, \nabla^\mu b \, \nabla_\mu b} , \cr
    &\nabla b = db + \mathrm{i} b\, d\phi .
    \end{split}
\end{equation}
The partition function of the IR fixed-point theory then is a ``zero-mode integral''
\begin{equation}\label{eq:Z-IR}
    Z_\ast^{\rm IR} := \int_{G/H} \!\!\! dg_H \, Z(g \cdot C_p) ,
\end{equation}
where $G = \mathrm{SO}(3)$, $H = \mathrm{SO}(2) \subset \mathrm{SO}(3)$ is the stabilizer of $C_p$, and $dg_H$ denotes the $G$-invariant measure on the symmetric space $G/H$ (a.k.a.\ the solid-angle form $\omega$ on $\mathrm{S}^2$). Thus we are suggesting that the infinite-volume Gibbs state is an $\mathrm{SO} (3)$-invariant integral (\ref{eq:Z-IR}) of extremal Gibbs states with partition function (\ref{eq:Z-Cp}).

To conclude, we have developed a fairly detailed if heuristic scenario as to how non-perturbative renormalization might take the $\mathrm{O}(3)$ nonlinear $\sigma$-model at $\theta = \pi$ to a fixed point with conformal symmetry. Far from being a mathematical proof, that scenario does satisfy a number of non-trivial consistency checks. However, the decisive physics question now is this: can our scenario explain the (so far unexplained) criticality of some other nonlinear $\sigma$-models at $\theta = \pi$ (as listed in Sect.\ \ref{sect:intro})? A positive answer will be given in a follow-up paper, where we are going to show that, in particular, the present scenario directly leads to the proposal of \cite{CFT-IQHT} for the infrared limit of Pruisken's model for the integer quantum Hall transition.

\bibliography{O3.bib}
\end{document}